\documentclass[11pt]{article}

\usepackage[utf8]{inputenc}
\usepackage[T1]{fontenc}
\usepackage[english]{babel}
\usepackage{amsmath,amssymb,amsthm}
\usepackage{booktabs}
\usepackage{multirow}
\usepackage{graphicx}
\usepackage{array}
\usepackage[margin=1in]{geometry}
\usepackage{tikz}
\usepackage{pgfplots}
\usepackage{algorithm}
\usepackage{algpseudocode}
\usepackage{caption}
\usepackage[numbers,sort&compress]{natbib}
\usepackage[colorlinks=true,linkcolor=blue,citecolor=blue,urlcolor=blue]{hyperref}
\usepackage{cleveref}

\pgfplotsset{compat=1.17}
\usetikzlibrary{positioning,arrows.meta,calc,fit,backgrounds}

\newtheorem{definition}{Definition}
\newtheorem{remark}{Remark}
\newtheorem{proposition}{Proposition}

\newcommand{\scc}{\mathrm{sc}}
\newcommand{\tw}{\mathrm{tw}}
\newcommand{\In}{\mathrm{in}}
\newcommand{\Out}{\mathrm{out}}
\newcommand{\yes}{\ensuremath{\checkmark}}
\newcommand{\no}{\ensuremath{\times}}

\tikzset{
  stage/.style={draw, rounded corners=2pt, align=center, font=\small,
                minimum height=8mm, inner xsep=4pt, fill=black!3},
  dec/.style={draw, rounded corners=2pt, align=center, font=\small,
              minimum height=8mm, inner xsep=4pt, fill=black!8},
  flow/.style={-{Latex[length=2mm]}, thick},
}

\title{\bfseries zLend: A Dual-Scope Cash-Flow Reconstruction Framework for
On-Chain Credit Underwriting}

\author{
{\large\bfseries Zeru AI}\\[0.9em]
\normalsize
\begin{tabular}{@{}c@{\hspace{2.4em}}c@{\hspace{2.4em}}c@{}}
  Girish G N & Ashutosh Sahoo & Akshay SP \\
  {\small\itshape Head of AI} &
  {\small\itshape Chief Executive Officer} &
  {\small\itshape Chief Product Officer} \\
  {\small\ttfamily girish@zeru.finance} &
  {\small\ttfamily ashutosh@zeru.finance} &
  {\small\ttfamily akshay@zeru.finance} \\
\end{tabular}\\[7mm]
\begin{tabular}{@{}c@{\hspace{2.4em}}c@{}}
  Gurukiran S & Dhanashekar Kandaswamy \\
  {\small\itshape Chief Technology Officer} &
  {\small\itshape Ohio State University, USA} \\
  {\small\ttfamily gurukiran@zeru.finance} & \\
\end{tabular}
}

\date{}

\begin{document}
\maketitle


\begin{abstract}
Decentralized lending lacks a credit bureau: a borrower's capacity to repay
must be inferred entirely from public on-chain activity, without income
verification or a liability record. This paper presents zLend, a deployed
cash-flow underwriting framework that reconstructs a wallet's daily balance
history from raw token transfers and derives short-duration
repayment-capacity signals from the reconstruction. The reconstruction is
performed twice per wallet, once restricted to a fixed stablecoin basket and
once over all fungible transfers, on the premise that a wallet's total token
holdings and its liquid, spendable balance are distinct quantities whose
conflation misprices risk. From each series we derive liquidity coverage
against a fixed loan size, cash-flow volatility and regularity, a
peak-to-trough drawdown and recovery statistic adapted from quantitative
finance, and a recurring-counterparty detector that identifies salary-like
payment cadence from transfer timing alone. The two views are then compared:
a wallet whose aggregate holdings are large but whose stablecoin reserve
rarely covers the loan size, or whose inflows are overwhelmingly
non-stablecoin, is flagged as a liquidity or flow mismatch irrespective of
total wealth. We specify the pipeline formally, document the golden-master
methodology used to verify a cross-language production migration to
numerical tolerance $10^{-9}$, and characterize the tier function's
parameter sensitivity using an independent reimplementation validated to
exact agreement (78 of 78 field assertions) against the deployed system's
reference fixtures. That analysis yields three findings: tier assignment is
governed predominantly by the reference loan size relative to which coverage
is defined, with four of six reference wallets changing tier across
$L \in [\$10,\ \$25{,}000]$; the drawdown and coverage criteria bind on
disjoint wallets, establishing that neither subsumes the other; and no
criterion in the tier rule is inert among the regimes tested. The same
analysis applied to the cross-scope mismatch flags finds genuine sensitivity
in two of three free parameters, and holds decisively across the tested
range on the third. This is a methodology paper: it specifies and justifies
the framework and
characterizes its decision surface. zLend is deployed in production, where
it informs real lending decisions through API integration with third-party
applications.
\end{abstract}

\section{Introduction}
\label{sec:intro}

Undercollateralized lending is the DeFi primitive least served by existing
on-chain infrastructure. An unsecured line of credit in traditional finance
rests on income verification, a liability record, and a bureau score
aggregating repayment behavior across lenders. A wallet requesting a small,
short-duration on-chain loan supplies none of these. It supplies a public,
permanent, but semantically flat transfer log in which income, savings,
speculation, and debt service carry no distinguishing label.

The obvious substitute is total wallet value. That substitution fails
structurally rather than marginally, because total token value conflates two
economically distinct quantities: capital a borrower could liquidate on
short notice to service a loan, and capital held in volatile or illiquid
form that the borrower may have no intention of spending. A wallet holding
\$200{,}000 in a governance token and \$30 in stablecoins is, with respect to
repaying a \$100 loan next week, indistinguishable from a wallet holding
\$130 outright; a total-value screen ranks the former three orders of
magnitude higher.

zLend resolves this by maintaining two parallel reconstructions of the same
wallet: a stablecoin-scoped balance history restricted to assets whose
dollar value requires no price oracle to interpret, and a total-wealth
history over all fungible transfers. Both are reconstructed by an identical
procedure and feed an identical family of liquidity, cash-flow, and risk
statistics. Divergence between the two views is then treated as signal
rather than noise, since it is precisely the wallets whose aggregate wealth
and liquid reserve disagree that a single-series model misprices.

\paragraph{Contributions.}
\begin{itemize}
\item A formal specification of dual-scope daily balance reconstruction
(\Cref{sec:recon}) from unordered transfer records, using a
cumulative-net-flow construction with a non-negativity offset that requires
no prior knowledge of the wallet's opening balance (\Cref{alg:recon}).

\item A signal family derived from the reconstructed series
(\Cref{sec:signals}): liquidity coverage against absolute and
wallet-relative thresholds, inflow regularity and recurring-counterparty
detection, maximum drawdown and recovery, single-day outflow concentration,
and normalized trend classification.

\item A cross-scope comparison layer (\Cref{sec:comparison}) converting
divergence between the two views into graded liquidity- and flow-mismatch
severities and a four-tier underwriting signal, specified completely in
\Cref{tab:tiers,tab:severity}.

\item A verified production implementation (\Cref{sec:impl}) whose
cross-language migration is checked against its reference to $10^{-9}$
numerical tolerance, including deliberate replication of summation,
variance-convention, and rounding-mode semantics.

\item A sensitivity and binding-constraint analysis (\Cref{sec:sensitivity})
conducted with an independent reimplementation validated to exact agreement
against the deployed system's reference fixtures, characterizing which
criteria in the tier rule actually determine outcomes.
\end{itemize}

The framework specified here is deployed in production and reaches real
borrowers today, integrated by third-party lending applications through the
API described in \Cref{sec:impl}.

\section{Related Work}
\label{sec:related}

\paragraph{Alternative and cash-flow credit scoring.}
The premise that behavioral data substitutes for an absent credit history
predates DeFi. Björkegren and Grissen show that mobile phone usage patterns
predict loan repayment comparably to a traditional bureau score in
populations without credit files \citep{bjorkegren2020}. Berg et al.\
demonstrate that elementary digital-footprint variables collected at
checkout match or exceed bureau-score predictive power for default
\citep{berg2020}. The operative signal in both is behavioral regularity
rather than self-reported financial position; the recurring-counterparty and
income-cadence detectors of \Cref{sec:signals} are its on-chain analogue.

\paragraph{DeFi lending risk.}
Undercollateralized lending inherits the risk vocabulary of collateralized
DeFi lending without its safety margin. Qin et al.\ characterize liquidation
events empirically across major protocols, finding they concentrate around
sharp price movements and are executed by a small set of sophisticated
searchers \citep{qin2021}. Gudgeon et al.\ analyze protocol mechanics and the
reflexive coupling between collateral value and solvency
\citep{gudgeon2020}; Aramonte et al.\ survey DeFi risk from a macroprudential
standpoint, emphasizing shock transmission through composability
\citep{aramonte2021}. zLend models no collateral position, but shares the
underlying concern that a point-in-time valuation poorly proxies a wallet's
condition under stress, which motivates the rolling-window drawdown
statistic of \Cref{sec:signals}.

\paragraph{Undercollateralized lending protocols.}
A small number of production DeFi protocols already extend credit below full
collateralization, each substituting a different form of off-chain trust for
on-chain collateral. Goldfinch routes underwriting through off-chain
\emph{backers} who stake first-loss junior capital against individual
borrowers following traditional due diligence, with a senior pool of passive
lenders bearing residual risk \citep{goldfinch2020}. Maple Finance delegates
credit assessment to \emph{pool delegates} --- vetted institutional managers
who conduct off-chain diligence, negotiate terms, and post first-loss
capital of their own \citep{maple2021}. TrueFi assesses creditworthiness
through token-holder staking and voting on individual loan requests, again
resting the underwriting judgment on off-chain information the protocol does
not itself observe \citep{truefi2020}. Centrifuge substitutes appraised
real-world collateral for behavioral signal entirely, tokenizing invoices and
other off-chain assets as loan backing \citep{centrifuge2018}. In each
design, the trust-critical judgment --- whether a borrower is creditworthy,
or an asset is worth what it claims --- is made by a human party outside the
protocol; the smart contract enforces an already-made decision rather than
deriving one. zLend instead computes its signal entirely from public
on-chain transfer history, with no off-chain underwriter, staked backer, or
appraised collateral in the loop, making it computable for any address with
sufficient transfer history rather than only the subset a human underwriter
has chosen to evaluate.

\paragraph{Drawdown as a risk measure.}
Maximum drawdown is standard in portfolio and strategy risk assessment, and
its statistical properties are characterized by Magdon-Ismail and Atiya
\citep{magdonismail2004}. We are not aware of prior work applying maximum
drawdown, computed over a reconstructed on-chain balance series, as an
underwriting criterion for unsecured credit.

\paragraph{Positioning.}
zLend is the credit-scoped counterpart to the zScore reputation lineage
\citep{zscore,deepreputation}. Where zScore establishes general-purpose
cross-protocol wallet reputation, zLend addresses the narrower question of
short-duration repayment capacity, over a different representation: a
reconstructed daily balance series rather than aggregate transaction
statistics.

\section{Problem Formulation and Balance Reconstruction}
\label{sec:recon}

\subsection{Setting}

Let $w$ denote a wallet and $\mathcal{T}_w$ its observed token transfers.
Each transfer $t \in \mathcal{T}_w$ carries a timestamp, a USD value
$v(t) \ge 0$, a token symbol $s(t)$, a transfer type, a direction
$\delta(t) \in \{\In, \Out\}$ relative to $w$, and a counterparty address
$\kappa(t)$. No balance is observed at any point: holdings must be inferred
from flows alone.

\subsection{Scopes}

Every wallet is analyzed under two scopes $\sigma \in \{\scc, \tw\}$ applied
to the same $\mathcal{T}_w$.

\begin{definition}[Scope filters]
\label{def:scopes}
Let $\mathrm{norm}(s)$ denote symbol normalization: uppercase, strip
whitespace, then map any symbol prefixed \texttt{USDC} to \texttt{USDC}, any
symbol prefixed \texttt{USDT} to \texttt{USDT}, and \texttt{USDBC} to
\texttt{USDC}. Then
\[
\mathcal{T}_w^{\scc} = \{t : \mathrm{type}(t)=\text{fungible} \wedge \mathrm{norm}(s(t)) \in \mathcal{S}\},
\qquad
\mathcal{T}_w^{\tw} = \{t : \mathrm{type}(t)=\text{fungible} \wedge v(t) > 0\},
\]
where $\mathcal{S}$ is the fixed 15-symbol stablecoin basket
\{USDC, USDT, DAI, FDUSD, USDE, PYUSD, GUSD, USDP, TUSD, BUSD, USDBC, USDB,
FRAX, USDS, LUSD\}.
\end{definition}

The $\scc$ scope approximates capital deployable toward repayment without
price risk over the loan term; $\tw$ captures the wallet's full economic
footprint. Normalization folds bridged and chain-specific variants of one
asset together, so a wallet holding USDC across several chains is not scored
as holding several distinct assets.

\subsection{Daily Aggregation and Derived Balance}

Let $d$ index UTC calendar days. For scope $\sigma$,
\begin{equation}
\In_\sigma(d) = \!\!\sum_{\substack{t \in \mathcal{T}_w^\sigma:\ \mathrm{day}(t)=d \\ \delta(t)=\In}}\!\! v(t),
\qquad
\Out_\sigma(d) = \!\!\sum_{\substack{t \in \mathcal{T}_w^\sigma:\ \mathrm{day}(t)=d \\ \delta(t)=\Out}}\!\! v(t),
\qquad
\nu_\sigma(d) = \In_\sigma(d) - \Out_\sigma(d).
\label{eq:daily-agg}
\end{equation}

Because no opening balance is observed, the balance series is built from
cumulative net flow and shifted by the minimum offset rendering it
non-negative:
\begin{equation}
C_\sigma(d) = \sum_{d' \le d} \nu_\sigma(d'),
\qquad
\omega_\sigma = \max\!\Big(0,\ -\min_{d} C_\sigma(d)\Big),
\qquad
B_\sigma(d) = C_\sigma(d) + \omega_\sigma .
\label{eq:balance}
\end{equation}
Here $C_\sigma$ and $\omega_\sigma$ range over the (generally sparse) set of
days on which scope $\sigma$ has at least one transfer; \Cref{alg:recon}
extends $B_\sigma$ to the full contiguous calendar.

\begin{definition}[Offset semantics]
\label{def:offset}
$\omega_\sigma$ is the smallest opening balance consistent with the observed
history never becoming negative. It is a lower bound implied by the data,
not an estimate of the wallet's true pre-history balance. Level-dependent
statistics inherit this bound; ratio- and variation-dependent statistics do
not.
\end{definition}

\subsection{Daily Spine}

Scope-level day sets are sparse and generally disjoint. Both are projected
onto a single contiguous calendar $\mathcal{D} = [d_{\min}, d_{\max}]$
spanning the earliest to latest transfer day across \emph{either} scope, so
that the two series are indexed identically and remain directly comparable.
\Cref{alg:recon} gives the full construction, including the opening-balance
series $O_\sigma$ required by \Cref{def:outflow}.

\begin{algorithm}[t]
\caption{Dual-scope daily balance reconstruction}
\label{alg:recon}
\begin{algorithmic}[1]
\Require transfers $\mathcal{T}_w$; stablecoin basket $\mathcal{S}$
\Ensure spine $\mathcal{D}$; per-scope series $B_\sigma, O_\sigma, \In_\sigma, \Out_\sigma$
\For{$\sigma \in \{\scc, \tw\}$}
  \State $\mathcal{T}_w^\sigma \gets$ filter $\mathcal{T}_w$ by \Cref{def:scopes}
  \State aggregate $\In_\sigma, \Out_\sigma$ per day by \Cref{eq:daily-agg}
    \Comment{compensated summation}
  \State $B_\sigma \gets$ cumulative net flow $+\ \omega_\sigma$ by \Cref{eq:balance}
\EndFor
\State $\mathcal{D} \gets [\,\min_\sigma \min \mathrm{days}(\sigma),\ \max_\sigma \max \mathrm{days}(\sigma)\,]$
\For{$\sigma \in \{\scc, \tw\}$}
  \State project $B_\sigma, \In_\sigma, \Out_\sigma$ onto $\mathcal{D}$; unobserved days take $\In=\Out=0$
  \State $B_\sigma \gets$ forward-fill over $\mathcal{D}$
    \Comment{a day without transfers retains the prior balance}
  \If{$B_\sigma$ is unobserved on all of $\mathcal{D}$}
    \State $B_\sigma(d) \gets C_\sigma(d)$ for all $d$
  \Else
    \State $d^\ast \gets$ first observed day;\quad $a \gets B_\sigma(d^\ast) - C_\sigma(d^\ast)$
    \State $B_\sigma(d) \gets C_\sigma(d) + a$ for every $d$ still unfilled
      \Comment{back-extend through the anchor}
  \EndIf
  \State $O_\sigma(d) \gets B_\sigma(d-1)$, falling back to $B_\sigma(d) - \nu_\sigma(d)$, then to $0$
\EndFor
\State \Return $\mathcal{D},\ \{B_\sigma, O_\sigma, \In_\sigma, \Out_\sigma\}$
\end{algorithmic}
\end{algorithm}

\begin{proposition}[Complexity]
\label{prop:complexity}
For $T = |\mathcal{T}_w|$ transfers spanning $D = |\mathcal{D}|$ days,
reconstruction costs $O(T \log T + D)$ time and $O(D)$ space. The statistic
suite adds $O(\sum_W W \log W)$ for order statistics over
$W \in \{30,60,90\}$, which is constant-bounded. The pipeline is therefore
effectively linear in transfer count.
\end{proposition}

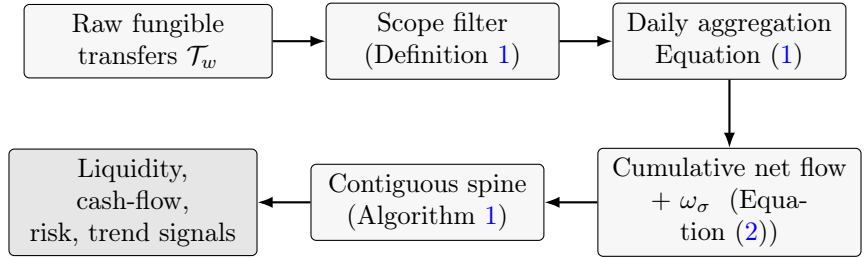
\begin{figure}[t]
\centering
\begin{tikzpicture}[node distance=6mm]
\node[stage, text width=30mm] (raw) {Raw fungible\\transfers $\mathcal{T}_w$};
\node[stage, right=7mm of raw, text width=28mm] (split) {Scope filter\\(\Cref{def:scopes})};
\node[stage, right=7mm of split, text width=28mm] (agg) {Daily aggregation\\\Cref{eq:daily-agg}};
\node[stage, below=9mm of agg, text width=32mm] (bal) {Cumulative net flow\\$+\ \omega_\sigma$\ \ (\Cref{eq:balance})};
\node[stage, left=7mm of bal, text width=28mm] (spine) {Contiguous spine\\(\Cref{alg:recon})};
\node[stage, left=7mm of spine, text width=30mm, fill=black!10] (sig) {Liquidity, cash-flow,\\risk, trend signals};

\draw[flow] (raw) -- (split);
\draw[flow] (split) -- (agg);
\draw[flow] (agg) -- (bal);
\draw[flow] (bal) -- (spine);
\draw[flow] (spine) -- (sig);
\end{tikzpicture}
\caption{Reconstruction pipeline, executed independently and identically for
the stablecoin and total-wealth scopes.}
\label{fig:pipeline}
\end{figure}

\Cref{fig:pipeline} summarizes the construction.

\section{Signal Derivation}
\label{sec:signals}

All statistics are computed per scope over rolling tail windows
$W \in \{30,60,90\}$ days, each taking the final $\min(W, |\mathcal{D}|)$
days of the spine, so that a wallet with less history than the nominal
window is scored over what is available rather than zero-padded.
\Cref{tab:params} lists every free parameter.

\begin{table}[t]
\centering
\caption{Pipeline parameters and deployed values.}
\label{tab:params}
\begin{tabular}{llr}
\toprule
Parameter & Role & Value \\
\midrule
$L$ & Reference loan size for coverage & \$100 \\
$\varepsilon$ & Zero-balance threshold & \$1 \\
$f_{\mathrm{abs}}$ & Absolute floor, outflow-concentration gate & \$5 \\
$f_{\mathrm{pct}}$ & Median-relative floor, same gate & $0.05$ \\
$k_{\mathrm{rec}}$ & Distinct days for counterparty recurrence & $3$ \\
$\tau$ & Flat-band tolerance, trend classification & $0.01$ \\
$\theta_{\mathrm{dyn}}$ & Dynamic balance threshold & 90-day median \\
\bottomrule
\end{tabular}
\end{table}

\subsection{Liquidity Coverage}

\begin{definition}[Coverage]
\label{def:coverage}
For loan size $L$,
\begin{equation}
\phi_\sigma^{(W)} = \frac{1}{|W|}\sum_{d \in W} \mathbb{1}\!\left[B_\sigma(d) \ge L\right]
\label{eq:coverage}
\end{equation}
is the fraction of window days on which the derived balance would have
covered the loan outright. A companion statistic replaces $L$ with
$\theta_{\mathrm{dyn}}$, the wallet's own 90-day median balance, measuring
coverage relative to the wallet's operating scale rather than an absolute
figure.
\end{definition}

\subsection{Cash-Flow Regularity}

Inflow frequency and the coefficient of variation of inflow amounts
characterize how often and how uniformly funds arrive. Counterparty
structure is then examined directly.

\begin{definition}[Recurrence and income cadence]
\label{def:recurrence}
A counterparty $\kappa$ is \emph{recurring} in window $W$ if it appears as
an inflow source on at least $k_{\mathrm{rec}}=3$ distinct days. Let
$g_1,\dots,g_m$ be the day-gaps between consecutive appearances. Then
$\kappa$ is \emph{income-like} if
\begin{equation}
\frac{\mathrm{sd}(g)}{\overline{g}} \le 0.5
\qquad\text{and}\qquad
5 \le \mathrm{median}(g) \le 45 ,
\label{eq:income}
\end{equation}
a band admitting weekly-to-monthly cadences while excluding both
high-frequency automated flows and coincidentally repeated transfers.
\end{definition}

\subsection{Drawdown, Depletion, and Outflow Concentration}

\begin{definition}[Maximum drawdown]
\label{def:mdd}
With running peak $P_\sigma(d) = \max_{d' \le d} B_\sigma(d')$, the drawdown
is $\mathrm{DD}_\sigma(d) = (P_\sigma(d) - B_\sigma(d))/P_\sigma(d)$,
evaluated only where $P_\sigma(d) > \varepsilon$, and
\begin{equation}
\mathrm{MDD}_\sigma^{(W)} = \max_{d \in W} \mathrm{DD}_\sigma(d).
\label{eq:mdd}
\end{equation}
Duration is the longest run of consecutive days below the prevailing peak;
recovery time is the number of days from the trough until $B_\sigma$ first
regains the pre-drawdown peak, undefined where no such day exists in the
window.
\end{definition}

\begin{definition}[Zero-balance event]
\label{def:zero}
$Z_\sigma(d) = \mathbb{1}\!\left[B_\sigma(d-1) \ge \varepsilon \wedge B_\sigma(d) < \varepsilon\right]$,
with $B_\sigma(d_{\min}-1) := 0$. Events are detected over the full spine
and then counted per window, so that an event adjacent to a window boundary
is neither double-counted nor lost.
\end{definition}

\begin{definition}[Single-day outflow concentration]
\label{def:outflow}
Let $\theta = \max\!\big(f_{\mathrm{abs}},\ f_{\mathrm{pct}}\cdot
\mathrm{median}_{d \in W} B_\sigma(d)\big)$. Over days with
$O_\sigma(d) \ge \theta$,
\begin{equation}
\Lambda_\sigma^{(W)} = \max_{d}\ \frac{\min\!\big(\Out_\sigma(d),\ O_\sigma(d)\big)}{O_\sigma(d)},
\label{eq:outflow}
\end{equation}
undefined if no day qualifies. The gate suppresses the degenerate case in
which a trivially small opening balance is fully withdrawn, which would
otherwise register as maximal-severity outflow.
\end{definition}

\subsection{Trend}

Net flow is regressed on day index by ordinary least squares within the
window and normalized by mean absolute balance, yielding a rate rather than
a currency amount:
\begin{equation}
\beta_\sigma^{(W)} =
\frac{\sum_i (i - \bar{i})\big(\nu_\sigma(d_i) - \bar{\nu}\big)}{\sum_i (i - \bar{i})^2}
\Big/ \max\!\big(|\bar{B}_\sigma^{(W)}|,\ 1\big).
\label{eq:trend}
\end{equation}
The label is \emph{increasing} if $\beta_\sigma^{(W)} > \tau$,
\emph{decreasing} if $\beta_\sigma^{(W)} < -\tau$, and \emph{flat}
otherwise. Comparing the 30-day slope $r$ against the lifetime slope $\ell$
gives a five-way transition label, evaluated in order and resolved on first
match:
\begin{equation}
\begin{aligned}
&\textit{accelerating inflow}  &&\text{if } r \ge \tau \wedge \ell \ge \tau \wedge r > \ell + \tau,\\
&\textit{accelerating outflow} &&\text{if } r \le -\tau \wedge \ell \le -\tau \wedge r < \ell - \tau,\\
&\textit{improving}            &&\text{if } r > \ell + \tau \wedge r > -\tau,\\
&\textit{weakening}            &&\text{if } r < \ell - \tau \wedge r < \tau,\\
&\textit{stable}               &&\text{otherwise.}
\end{aligned}
\label{eq:transition}
\end{equation}

\subsection{Data-Quality Flags}

Two conditions qualify every record. A wallet is flagged
\emph{insufficient-history} when $|\mathcal{D}| < 90$. Separately, let the
counterparty coverage of scope $\sigma$ be the fraction of 90-day inflow
days carrying at least one resolved counterparty address; if either scope
falls below $0.8$, the record is flagged
\emph{insufficient-counterparty-data}, since \Cref{def:recurrence} is then
evaluated over an incomplete graph and recurrence may be understated. This
90-day quality threshold is diagnostic and distinct from the 30-day
\emph{insufficient} tier of \Cref{tab:tiers}, which determines the primary
credit signal itself rather than flagging the confidence of a signal that
was still assigned.

\section{Cross-Scope Comparison and Tier Assignment}
\label{sec:comparison}

The signals of \Cref{sec:signals} are computed identically for both scopes.
Their comparison is where the framework's central design decision operates.

\subsection{Mismatch Detection}

\begin{definition}[Liquidity mismatch]
\label{def:liqmis}
With $\widetilde{B}^{(30)}$ the 30-day median balance,
\begin{equation}
\mathrm{LM} = \mathbb{1}\!\left[\,
\widetilde{B}_\tw^{(30)} \ge 3\,\widetilde{B}_\scc^{(30)}
\ \wedge\ \phi_\scc^{(30)} < 0.5 \,\right].
\label{eq:liqmismatch}
\end{equation}
\end{definition}

\begin{definition}[Flow mismatch]
\label{def:flowmis}
With $\rho = \In_\scc^{(30)} / \In_\tw^{(30)}$ the stablecoin share of
trailing inflow,
\begin{equation}
\mathrm{FM} = \mathbb{1}\!\left[\, \In_\tw^{(30)} > 0 \ \wedge\ \rho < 0.25 \,\right].
\label{eq:flowmismatch}
\end{equation}
\end{definition}

Both flags carry graded severity (\Cref{tab:severity}), so that a wallet
marginally past a threshold is not treated identically to one exceeding it
by orders of magnitude. A four-way trend-alignment label
(\emph{aligned-positive}, \emph{aligned-flat}, \emph{aligned-negative},
\emph{divergent}) records whether the two scopes' 30-day trend labels agree.

\begin{table}[t]
\centering
\caption{Mismatch severity ladders, resolved most-severe-first: a wallet
satisfying the High condition is High regardless of whether it also
satisfies Low or Medium.
$r = \widetilde{B}_\tw^{(30)}/\widetilde{B}_\scc^{(30)}$ is the balance
ratio; $\rho$ is the stablecoin inflow share.}
\label{tab:severity}
\begin{tabular}{lll}
\toprule
Severity & Liquidity mismatch & Flow mismatch \\
\midrule
None   & flag not raised & flag not raised \\
Low    & $r < 4$ and $\phi_\scc^{(30)} \ge 0.35$ & $\rho \ge 0.2$ \\
Medium & $r \ge 4$ or $\phi_\scc^{(30)} < 0.35$ & $\rho < 0.2$ \\
High   & $r \ge 6$ or $\phi_\scc^{(30)} < 0.2$ & $\rho < 0.1$ \\
\bottomrule
\end{tabular}
\end{table}

\subsection{Tier Assignment}

The stablecoin scope alone determines the primary credit signal; the
total-wealth scope yields a parallel context signal that qualifies but never
overrides it. \Cref{tab:tiers} specifies both.

\begin{table}[t]
\centering
\caption{Tier definitions. Conditions within a row are conjunctive except
where noted; rows are evaluated top to bottom and resolved on first match.}
\label{tab:tiers}
\begin{tabular}{llcccl}
\toprule
Signal & Tier & $\phi^{(30)}$ & Zero events$^{(90)}$ & $\mathrm{MDD}^{(90)}$ & Trend$^{(30)}$ \\
\midrule
\multirow{4}{*}{Primary ($\scc$)}
 & Insufficient & \multicolumn{4}{l}{$|\mathcal{D}| < 30$} \\
 & Strong   & $\ge 0.80$ & $=0$    & $\le 0.35$ & $\ne$ decreasing \\
 & Moderate & $\ge 0.50$ & $\le 1$ & $\le 0.65$ & --- \\
 & Weak     & \multicolumn{4}{l}{otherwise} \\
\midrule
\multirow{3}{*}{Context ($\tw$)}
 & Supportive & $\ge 0.80$ & --- & $\le 0.50$ & $\ne$ decreasing \\
 & Cautionary & \multicolumn{4}{l}{$\phi < 0.50$ \emph{or} zero events $> 0$ \emph{or} $\mathrm{MDD} > 0.75$ (disjunctive)} \\
 & Neutral    & \multicolumn{4}{l}{otherwise} \\
\bottomrule
\end{tabular}
\end{table}

\begin{remark}[Conjunctivity]
\label{rem:conjunctive}
Primary-tier criteria are conjunctive and non-substitutable: a wallet
failing any one cannot reach that tier however strongly it satisfies the
others. This is deliberate. Coverage measures whether a wallet is
\emph{usually} solvent against the loan; drawdown measures whether it has
\emph{recently demonstrated} that it can lose nearly everything. A wallet
can satisfy the first perfectly while failing the second, and
\Cref{sec:examples} exhibits such a case. The context signal's cautionary
branch is by contrast disjunctive, so that any single severe indicator on
the total-wealth view withdraws support.
\end{remark}

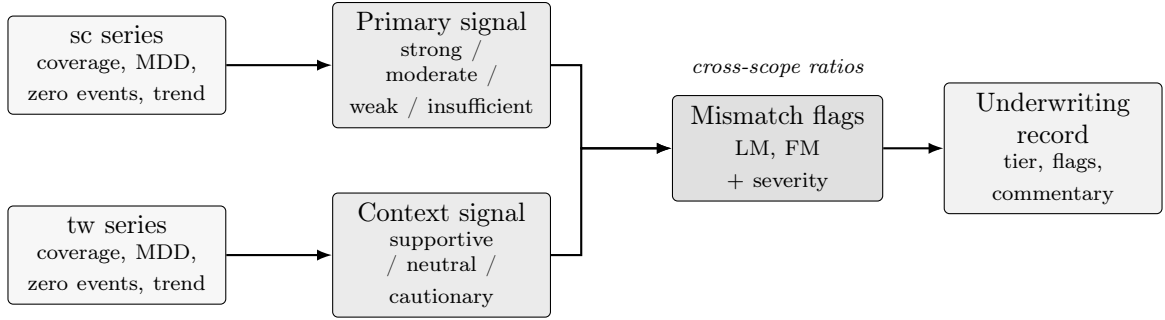
\begin{figure}[t]
\centering
\begin{tikzpicture}[node distance=5mm]
\node[stage, text width=26mm] (sc) {$\scc$ series\\{\scriptsize coverage, MDD,\\zero events, trend}};
\node[stage, below=12mm of sc, text width=26mm] (tw) {$\tw$ series\\{\scriptsize coverage, MDD,\\zero events, trend}};

\node[dec, right=14mm of sc, text width=26mm] (prim) {Primary signal\\{\scriptsize strong / moderate /\\weak / insufficient}};
\node[dec, right=14mm of tw, text width=26mm] (ctx) {Context signal\\{\scriptsize supportive / neutral /\\cautionary}};

\node[dec, right=16mm of prim, text width=25mm, fill=black!12, yshift=-11mm] (mis)
  {Mismatch flags\\{\scriptsize LM, FM $+$ severity}};
\node[stage, right=8mm of mis, text width=26mm, fill=black!5] (out)
  {Underwriting record\\{\scriptsize tier, flags,\\commentary}};

\draw[flow] (sc) -- (prim);
\draw[flow] (tw) -- (ctx);
\draw[flow] (prim.east) -- ++(4mm,0) |- (mis.west);
\draw[flow] (ctx.east) -- ++(4mm,0) |- (mis.west);
\draw[flow] (mis) -- (out);
\node[font=\scriptsize\itshape, above=1mm of mis] {cross-scope ratios};
\end{tikzpicture}
\caption{Decision flow. The two scopes are scored independently into
parallel signals; mismatch flags are computed from cross-scope ratios and
qualify, but never override, the stablecoin-derived primary tier.}
\label{fig:decision}
\end{figure}

\Cref{fig:decision} traces the full decision path.

\section{Implementation and Verification}
\label{sec:impl}

zLend is deployed as an in-process module of a production wallet-scoring
service. Each evaluation writes a record of 166 columns into an append-only
table partitioned monthly under a 90-day retention policy: per-scope
liquidity, cash-flow, and risk metrics (69 fields per scope), six
cross-scope ratios, the mismatch and tier fields, data-quality flags, and
four per-stage latency instrumentation columns.

The pipeline was implemented in Python and subsequently ported to TypeScript
during a service consolidation. Because its outputs inform a financial
decision, the port was verified rather than reviewed. A golden-master suite
generates reference output from the live Python implementation over
scenarios constructed to exercise each branch, and the TypeScript
implementation is asserted to reproduce every numeric field within $10^{-9}$
absolute tolerance and every categorical field exactly. Meeting that bound
required replicating numerical semantics that a straightforward
reimplementation would not preserve:

\begin{itemize}
\item daily aggregation uses compensated (Kahan) summation, matching the
reference implementation's dataframe groupby rather than naive accumulation;
\item dispersion is computed as a population statistic (denominator $n$),
matching the reference convention rather than the sample default;
\item output rounding replicates round-half-to-even at ten decimal places
rather than the target language's native mode;
\item ratios with denominator below $10^{-12}$ in absolute value yield
$\mathrm{NaN}$, propagated to the API boundary as an explicit null rather
than coerced to zero, preserving the distinction between undefined and zero.
\end{itemize}

This level of numerical fidelity is what makes zLend's production deployment
possible: third-party lending applications integrate against the API
described above, and each depends on the service returning exactly the
values the specification defines, call after call.

\section{Worked Examples}
\label{sec:examples}

The wallets below are drawn from the deployed system's verification
fixtures. Each is a synthetic transfer history constructed to exercise a
specific regime; every reported value is verified output of the
specification, reproduced independently for this paper and checked against
the reference fixtures. \Cref{tab:examples} summarizes all six.

\begin{table}[t]
\centering
\caption{Reference wallets, grouped by primary tier. All values are
stablecoin-scope.}
\label{tab:examples}
\begin{tabular}{lccccl}
\toprule
Wallet regime & $|\mathcal{D}|$ & $\phi_\scc^{(30)}$ & $\mathrm{MDD}_\scc^{(90)}$ & Zero events & Primary tier \\
\midrule
Steady, low volume         & 86 & $1.00$ & $0.020$ & $0$ & Strong \\
Recurring income           & 95 & $1.00$ & $0.092$ & $0$ & Strong \\
Short history              & 36 & $1.00$ & $0.250$ & $0$ & Strong \\
Severe drawdown, recovered & 76 & $1.00$ & $0.950$ & $0$ & Weak \\
Repeated depletion         & 79 & $0.40$ & $1.000$ & $5$ & Weak \\
Thin stablecoin reserve    & 80 & $0.00$ & $0.250$ & $0$ & Weak \\
\bottomrule
\end{tabular}
\end{table}

The remaining three cases exercise regimes not covered by the two detailed
examples below. The steady, low-volume wallet is the unremarkable baseline
the strong tier is designed to admit: full coverage, negligible drawdown, no
exceptional signal on either scope. The recurring-income wallet exercises
\Cref{def:recurrence} directly: it receives a fixed \$2{,}500 stablecoin
payment on a seven-day cadence that the income-cadence detector correctly
flags as income-like, and its balance dips no more than $9.2\%$ below peak,
comfortably within the strong-tier drawdown bound. The short-history wallet
has only 36 days of observed activity --- above the 30-day insufficiency
floor of \Cref{tab:tiers} but below the 90-day data-quality threshold of
\Cref{sec:signals} --- and is admitted to the strong tier on the strength of
its available history alone, exercising the framework's decision to score
partial history rather than withhold judgment.

\subsection{Coverage without solvency stability}

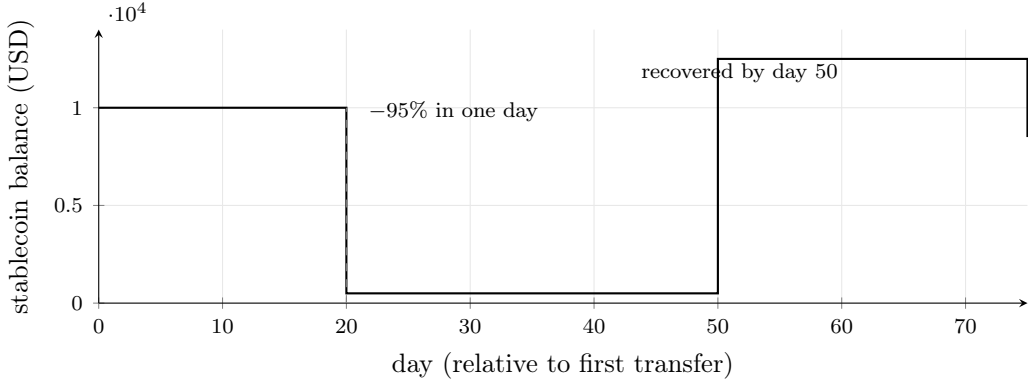
\begin{figure}[t]
\centering
\begin{tikzpicture}
\begin{axis}[
  width=0.84\textwidth, height=5.2cm,
  xlabel={day (relative to first transfer)},
  ylabel={stablecoin balance (USD)},
  xmin=0, xmax=75, ymin=0, ymax=14000,
  tick label style={font=\scriptsize}, label style={font=\small},
  grid=major, grid style={gray!18}, axis lines=left,
]
\addplot[thick, black, const plot] coordinates {
(0,10000)(19,10000)(20,500)(39,500)(40,500)(49,500)(50,12500)(74,12500)(75,8500)};
\draw[dashed, gray] (axis cs:20,0) -- (axis cs:20,10000);
\node[font=\scriptsize, anchor=south west] at (axis cs:21,8800) {$-95\%$ in one day};
\node[font=\scriptsize, anchor=south west] at (axis cs:43,10800) {recovered by day 50};
\end{axis}
\end{tikzpicture}
\caption{Reconstructed stablecoin balance for the drawdown wallet. Coverage
is $100\%$ across the trailing 30 days despite a 90-day maximum drawdown of
$0.95$.}
\label{fig:drawdown}
\end{figure}

\Cref{fig:drawdown} plots the reconstructed stablecoin balance of the
drawdown wallet. It holds \$10{,}000 for nineteen days, loses 95\% of that
in a single day, remains near \$500 for three weeks, then recovers past its
original level by day 50. Its coverage is $\phi_\scc^{(30)} = 1.00$: on
every one of the trailing thirty days the balance exceeded the \$100 loan
size, and a coverage-only screen would rate its liquidity perfect. Its
90-day maximum drawdown is $0.95$, well past the $0.35$ bound of
\Cref{tab:tiers}, and the primary signal is therefore \emph{weak}. This is
\Cref{rem:conjunctive} operating as designed.

\subsection{The mismatch the dual-scope design targets}

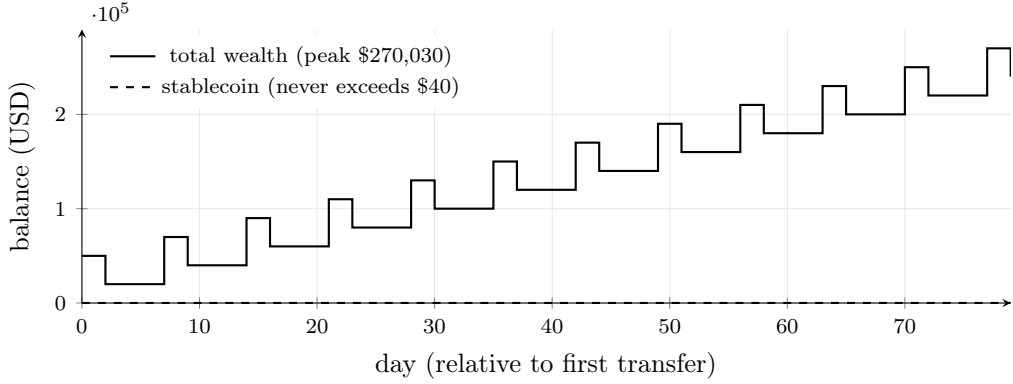
\begin{figure}[t]
\centering
\begin{tikzpicture}
\begin{axis}[
  width=0.84\textwidth, height=5.2cm,
  xlabel={day (relative to first transfer)},
  ylabel={balance (USD)},
  xmin=0, xmax=79, ymin=0, ymax=290000,
  tick label style={font=\scriptsize}, label style={font=\small},
  legend style={font=\scriptsize, at={(0.02,0.98)}, anchor=north west, draw=none, fill=white},
  grid=major, grid style={gray!18}, axis lines=left,
]
\addplot[thick, black, const plot] coordinates {
(0,50000)(1,50000)(2,20000)(6,20000)(7,70000)(8,70000)(9,40000)(13,40000)
(14,90040)(15,90040)(16,60040)(20,60040)(21,110040)(22,110040)(23,80040)(27,80040)
(28,130040)(29,130040)(30,100040)(34,100040)(35,150040)(36,150040)(37,120040)(41,120040)
(42,170040)(43,170040)(44,140040)(48,140040)(49,190040)(50,190040)(51,160040)(55,160040)
(56,210040)(57,210040)(58,180040)(59,180040)(60,180030)(62,180030)(63,230030)(64,230030)
(65,200030)(69,200030)(70,250030)(71,250030)(72,220030)(76,220030)(77,270030)(78,270030)(79,240030)};
\addlegendentry{total wealth (peak \$270{,}030)}
\addplot[thick, black, dashed, const plot] coordinates {(0,0)(9,0)(10,40)(79,40)};
\addlegendentry{stablecoin (never exceeds \$40)}
\end{axis}
\end{tikzpicture}
\caption{Both reconstructions for one wallet over the same 80-day history.
The views diverge by roughly four orders of magnitude throughout.}
\label{fig:mismatch}
\end{figure}

\Cref{fig:mismatch} plots both reconstructions for a wallet whose
total-wealth balance climbs from \$50{,}000 to \$270{,}030 while its
stablecoin balance never exceeds \$40. Stablecoin coverage is
$\phi_\scc^{(30)} = 0.00$: on no day in the trailing thirty could it have
covered a \$100 loan from stablecoins. The liquidity mismatch of
\Cref{eq:liqmismatch} fires at high severity, the balance ratio exceeding
$6{,}000$; the flow mismatch of \Cref{eq:flowmismatch} also fires at high
severity, stablecoin inflow being $0\%$ of trailing total inflow. A
total-value screen would rank this wallet among the wealthiest of any
comparable population. The two screens disagree by roughly four orders of
magnitude, and that disagreement is the signal.

\section{Sensitivity and Binding-Constraint Analysis}
\label{sec:sensitivity}

The tier rule of \Cref{tab:tiers} is specified by hand. This section
characterizes how far its output depends on that specification. Results were
produced by an independent reimplementation of the pipeline, validated
against the deployed system's reference fixtures to exact agreement on every
checked field (78 of 78 assertions at tolerance $10^{-9}$), then re-run
under perturbed parameters.

\begin{remark}[Interpretation]
\label{rem:sensinterp}
This analysis characterizes the \emph{decision surface of the tier
function}, not the distribution of a wallet population. The six reference
wallets were constructed to span the pipeline's designed operating regimes,
so the statements below describe how the rule responds to its parameters and
must not be read as frequencies in live traffic.
\end{remark}

\subsection{Loan-size sensitivity}

Coverage is defined relative to $L$, so $L$ propagates into every tier
decision. \Cref{tab:sens-loan} sweeps it across three orders of magnitude.

\begin{table}[t]
\centering
\caption{Primary tier as a function of reference loan size $L$.
S~$=$~strong, M~$=$~moderate, W~$=$~weak.}
\label{tab:sens-loan}
\begin{tabular}{lcccccccccc}
\toprule
Wallet regime & \$10 & \$50 & \$100 & \$250 & \$500 & \$1k & \$2.5k & \$5k & \$10k & \$25k \\
\midrule
Recurring income        & S & S & S & S & S & S & S & S & S & W \\
Short history           & S & S & S & S & S & S & M & W & W & W \\
Severe drawdown         & W & W & W & W & W & W & W & W & W & W \\
Repeated depletion      & W & W & W & W & W & W & W & W & W & W \\
Thin stablecoin reserve & S & W & W & W & W & W & W & W & W & W \\
Steady, low volume      & S & S & S & S & S & S & S & S & W & W \\
\bottomrule
\end{tabular}
\end{table}

Four of six wallets change tier across $L \in [\$10,\ \$25{,}000]$, and the
transitions are informative rather than arbitrary. The thin-stablecoin
wallet is rated \emph{strong} at $L=\$10$ and \emph{weak} at $L=\$50$
because its stablecoin balance sits near \$30--40: it genuinely can cover a
\$10 loan and genuinely cannot cover a \$50 one, and the tier tracks that
fact. The two wallets that never change tier are those excluded by drawdown
and depletion criteria, which are scale-free and therefore independent of
$L$.

The practical consequence is that $L$ is not a tuning constant but a
statement of product scope. A deployment lending \$100 and one lending
\$10{,}000 are not running the same classifier at different thresholds; they
are asking different questions, and one wallet may legitimately answer one
affirmatively and the other negatively.

\subsection{Drawdown-threshold sensitivity}

\Cref{tab:sens-dd} varies the strong-tier drawdown bound in isolation.

\begin{table}[t]
\centering
\caption{Primary tier as a function of the strong-tier drawdown bound
(default $0.35$).}
\label{tab:sens-dd}
\begin{tabular}{lccccccc}
\toprule
Wallet regime & $0.15$ & $0.25$ & $0.35$ & $0.50$ & $0.65$ & $0.80$ & $0.95$ \\
\midrule
Recurring income        & S & S & S & S & S & S & S \\
Short history           & M & S & S & S & S & S & S \\
Severe drawdown         & W & W & W & W & W & W & S \\
Repeated depletion      & W & W & W & W & W & W & W \\
Thin stablecoin reserve & W & W & W & W & W & W & W \\
Steady, low volume      & S & S & S & S & S & S & S \\
\bottomrule
\end{tabular}
\end{table}

Tier assignment is markedly more stable in this parameter than in $L$: four
of six wallets are invariant across the full range. The bound is
consequential only near a wallet's own realized drawdown, which is the
expected behavior of a threshold on a continuous statistic, and the deployed
default of $0.35$ sits adjacent to no wallet's value. The severe-drawdown
wallet is admitted only at a bound of $0.95$, at which point the criterion
excludes nothing.

\subsection{Which criteria bind}

A tier rule may contain criteria that never determine an outcome.
\Cref{tab:binding} records which strong-tier conditions fail, per wallet.

\begin{table}[t]
\centering
\caption{Strong-tier criterion satisfaction under deployed parameters.
\yes~$=$~satisfied, \no~$=$~fails.}
\label{tab:binding}
\begin{tabular}{lccccl}
\toprule
Wallet regime & $\phi \ge 0.8$ & zero $=0$ & $\mathrm{MDD} \le 0.35$ & trend & Tier \\
\midrule
Recurring income        & \yes & \yes & \yes & flat & Strong \\
Short history           & \yes & \yes & \yes & flat & Strong \\
Steady, low volume      & \yes & \yes & \yes & flat & Strong \\
Severe drawdown         & \yes & \yes & \no  & flat & Weak \\
Repeated depletion      & \no  & \no  & \no  & decreasing & Weak \\
Thin stablecoin reserve & \no  & \yes & \yes & flat & Weak \\
\midrule
Failures across the set & 2 & 1 & 2 & 1 & \\
\bottomrule
\end{tabular}
\end{table}

Every criterion is decisive for at least one wallet, and --- more
informatively --- the coverage and drawdown criteria bind on \emph{disjoint}
wallets. The severe-drawdown wallet fails on drawdown alone while satisfying
coverage perfectly; the thin-stablecoin wallet fails on coverage alone while
satisfying drawdown comfortably. Over this set the two statistics are
therefore not proxies for one another, which is the design premise of
\Cref{rem:conjunctive} and the principal justification for retaining both.
This establishes non-redundancy over the regimes tested, not over an
arbitrary wallet population.

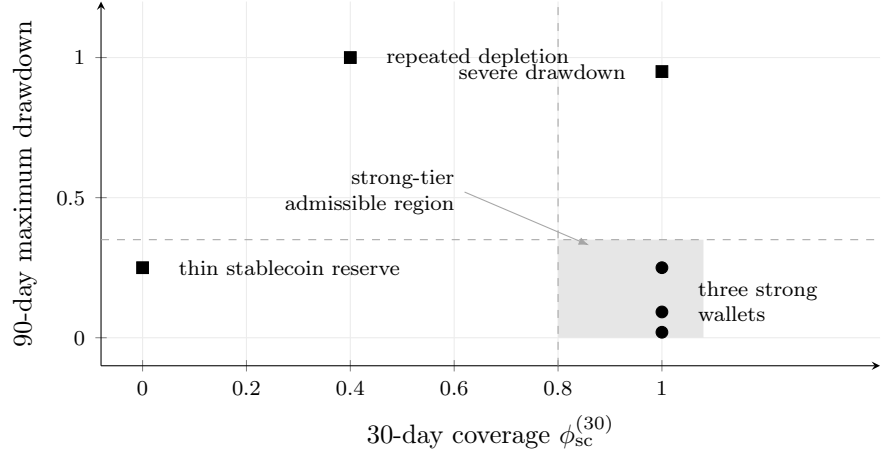
\begin{figure}[t]
\centering
\begin{tikzpicture}
\begin{axis}[
  width=0.72\textwidth, height=6.4cm,
  xlabel={30-day coverage $\phi_\scc^{(30)}$},
  ylabel={90-day maximum drawdown},
  xmin=-0.08, xmax=1.42, ymin=-0.10, ymax=1.20,
  xtick={0,0.2,0.4,0.6,0.8,1.0},
  tick label style={font=\scriptsize}, label style={font=\small},
  grid=major, grid style={gray!15}, axis lines=left,
]
\addplot[draw=none, fill=black!10, forget plot] coordinates
  {(0.8,0) (1.08,0) (1.08,0.35) (0.8,0.35)} \closedcycle;
\draw[dashed, gray!70] (axis cs:0.8,-0.10) -- (axis cs:0.8,1.20);
\draw[dashed, gray!70] (axis cs:-0.08,0.35) -- (axis cs:1.42,0.35);

\addplot[only marks, mark=*, mark size=2.2pt, black] coordinates {
(1.00,0.0196)(1.00,0.0920)(1.00,0.2500)};
\addplot[only marks, mark=square*, mark size=2.2pt, black] coordinates {
(1.00,0.9500)(0.40,1.0000)(0.00,0.2500)};

\node[font=\scriptsize, anchor=east, align=right] (reglab) at (axis cs:0.62,0.52)
  {strong-tier\\admissible region};
\draw[-{Latex[length=1.4mm]}, gray!70] (reglab.east) -- (axis cs:0.86,0.33);

\node[font=\scriptsize, anchor=west, align=left] at (axis cs:1.05,0.13)
  {three strong\\wallets};
\node[font=\scriptsize, anchor=east] at (axis cs:0.95,0.95) {severe drawdown};
\node[font=\scriptsize, anchor=west] at (axis cs:0.45,1.00) {repeated depletion};
\node[font=\scriptsize, anchor=west] at (axis cs:0.05,0.25) {thin stablecoin reserve};
\end{axis}
\end{tikzpicture}
\caption{Reference wallets in the (coverage, drawdown) plane. Circles denote
strong-tier wallets, squares weak. The severe-drawdown and thin-stablecoin
wallets fail on orthogonal criteria, so neither statistic subsumes the
other.}
\label{fig:plane}
\end{figure}

\Cref{fig:plane} makes the point geometrically: the strong-tier admissible
region is the lower-right rectangle, the three admitted wallets lie inside
it, and the two single-criterion failures lie outside it along orthogonal
axes.

\subsection{Mismatch-threshold sensitivity}
\label{sec:mismatch-sens}

The primary-tier criteria have now been stress-tested; the cross-scope
mismatch flags of \Cref{def:liqmis,def:flowmis} have not, despite motivating
the paper's central example (\Cref{sec:examples}). We extend the validated
reimplementation with the mismatch and severity rules of
\Cref{eq:liqmismatch,eq:flowmismatch} and \Cref{tab:severity}, re-verify it
against independently hand-derived values for three of the six reference
wallets (exact agreement on both flag and severity), and sweep each free
parameter in turn.

\textbf{The balance-ratio threshold (default $3\times$) holds decisively
across the tested range.} Sweeping it from $1.5\times$ to $10\times$
produces zero flag transitions: the two wallets that satisfy
\Cref{eq:liqmismatch} do so by wide margins, at a ratio of approximately
$6{,}668\times$ for the thin-stablecoin-reserve wallet, and at the
$10^{-9}$ numerical floor used in \Cref{eq:liqmismatch} to avoid division by
zero for the repeated-depletion wallet, whose 30-day stablecoin median is
exactly zero. The reference fixtures were designed primarily to probe the
tier's coverage and drawdown dimensions; extending them to more finely
probe the mismatch rule's ratio dimension is a natural next step.

The two remaining free parameters \emph{are} sensitive.
\Cref{tab:sens-covmis} sweeps the liquidity-mismatch coverage threshold
(default $0.5$): the repeated-depletion wallet, whose 30-day coverage is
$0.40$, is unflagged below a threshold of $0.5$ and flagged at or above it
--- a genuine boundary crossing of the same kind as \Cref{tab:sens-loan}.
\Cref{tab:sens-flowmis} sweeps the flow-mismatch share threshold (default
$0.25$): three wallets whose stablecoin inflow share falls between $0.35$
and $0.71$ are unflagged at the deployed default but would be flagged under
a more conservative one, indicating the default is permissive relative to
the range of stablecoin-inflow shares this fixture set exhibits.

\begin{table}[t]
\centering
\caption{Liquidity-mismatch flag under the coverage-threshold sweep (default
$0.5$, \Cref{eq:liqmismatch}). \no\ = not flagged.}
\label{tab:sens-covmis}
\begin{tabular}{lccccc}
\toprule
Wallet regime & $0.20$ & $0.35$ & $0.50$ & $0.65$ & $0.80$ \\
\midrule
Recurring income        & \no & \no & \no & \no & \no \\
Short history           & \no & \no & \no & \no & \no \\
Severe drawdown         & \no & \no & \no & \no & \no \\
Repeated depletion      & \no & \no & High & High & High \\
Thin stablecoin reserve & High & High & High & High & High \\
Steady, low volume      & \no & \no & \no & \no & \no \\
\bottomrule
\end{tabular}
\end{table}

\begin{table}[t]
\centering
\caption{Flow-mismatch flag under the share-threshold sweep (default
$0.25$, \Cref{eq:flowmismatch}). Severity is resolved most-severe-first: a
wallet satisfying the High condition of \Cref{tab:severity} is High
regardless of whether it also satisfies Low or Medium.}
\label{tab:sens-flowmis}
\begin{tabular}{lcccccccc}
\toprule
Wallet regime & $0.05$ & $0.10$ & $0.15$ & $0.20$ & $0.25$ & $0.35$ & $0.50$ & $0.75$ \\
\midrule
Recurring income        & \no & \no & \no & \no & \no & Low & Low & Low \\
Short history           & \no & \no & \no & \no & \no & \no & \no & \no \\
Severe drawdown         & \no & \no & \no & \no & \no & \no & \no & Low \\
Repeated depletion      & \no & \no & \no & \no & \no & \no & \no & \no \\
Thin stablecoin reserve & High & High & High & High & High & High & High & High \\
Steady, low volume      & \no & \no & \no & \no & \no & \no & \no & Low \\
\bottomrule
\end{tabular}
\end{table}

\section{Conclusion}
\label{sec:conclusion}

We presented zLend, a deployed framework that reconstructs a wallet's daily
balance history under two scopes --- stablecoin-only and total-wealth ---
and derives short-duration underwriting signals from the reconstruction:
liquidity coverage, cash-flow regularity, a drawdown and recovery statistic
adapted from quantitative finance, and a cross-scope comparison treating
divergence between aggregate holdings and liquid reserves as the
underwriting signal itself. The specification is given in full, and the
production implementation is verified against its reference to $10^{-9}$
numerical tolerance.

Sensitivity analysis over a validated reimplementation establishes that tier
assignment is governed principally by the reference loan size --- properly
understood as a statement of product scope rather than a tuning constant ---
and that the coverage and drawdown criteria bind on disjoint wallets,
confirming that neither subsumes the other. This rigor is what the
framework's production deployment rests on: zLend runs today inside real
lending applications, reached through the API of \Cref{sec:impl}, turning a
wallet's public transfer history into an underwriting signal at the moment a
loan decision is made.


\end{document}